\documentclass[prl,twocolumn,showpacs,amsmath,amssymb,superscriptaddress]{revtex4-1}
\usepackage{graphicx}
\usepackage{ulem}
\usepackage{graphicx}
\usepackage{dcolumn}
\usepackage{bm}
\usepackage{mathrsfs}
\usepackage{subfigure}
\usepackage{natbib}
\usepackage{color}
\usepackage{amssymb}
\usepackage{amsmath}

\begin{document}

\title{Critical superconductors}

\author{A. Vagov}
\affiliation{Institut f\"{u}r Theoretische Physik III, Bayreuth
Universit\"{a}t, Bayreuth 95440, Germany}

\author{A. A. Shanenko}
\affiliation{Departement Fysica, Universiteit Antwerpen, Groenenborgerlaan
171, B-2020 Antwerpen, Belgium} \affiliation{Departamento de F\'isica,
Universidade Federal de Pernambuco, 50670-901 Recife, PE, Brazil}

\author{M. V. Milo\v{s}evi\'{c}}
\affiliation{Departement Fysica, Universiteit Antwerpen, Groenenborgerlaan
171, B-2020 Antwerpen, Belgium}

\author{V. M. Axt}
\affiliation{Institut f\"{u}r Theoretische Physik III, Bayreuth
Universit\"{a}t, Bayreuth 95440, Germany}

\author{V. M. Vinokur}
\affiliation{Materials Science Division, Argonne National Laboratory, USA}

\author{F. M. Peeters}
\affiliation{Departement Fysica, Universiteit Antwerpen, Groenenborgerlaan
171, B-2020 Antwerpen, Belgium}

\date{\today}
\begin{abstract}
Bogomolnyi critical point, originally introduced in string theories, is
fundamental to superconductivity. At the critical temperature $T_c$ it
marks the sharp border between ideally diamagnetic bulk type-I
superconductors and type-II ones that can host vortices, while itself it
harbors infinitely degenerate distributions of magnetic flux in the
superconducting state. Here we show that below $T_c$ the physics of the
Bogomolnyi critical point projects onto a wider range of microscopic
parameters, even extremely wide for multiband superconductors. In this
critical range, the degeneracy of the superconducting state at $T_c$ is
lifted into a sequence of novel topological equilibria and the customary
understanding of superconducting phenomena does not suffice. As a radical
departure from traditional views, we introduce the paradigm of {\it
critical superconductors}, discuss their distinct magnetic properties,
advocate their subdivision in terms of possible intermediate states, and
demonstrate direct relevance of this paradigm to superconducting materials
of prime interest today.
\end{abstract}
\maketitle

Following textbooks \cite{abr,degen,kett}, superconductors are classified
by the relation between the magnetic penetration depth $\lambda$ and the
coherence length $\xi$. According to the Ginzburg-Landau (GL) theory
\cite{GL} types I and II of superconductivity interchange when the GL
parameter $\kappa=\lambda/\xi$ crosses the critical value $\kappa_0 =
1/\sqrt{2}$. However, detailed experimental studies as well as the
microscopic theory revealed that the interchange actually occurs in a
rather narrow but finite interval around $\kappa_0$~\cite{typeII/1_expA,
typeII/1_expB,typeII/1_theor_J,type_II/1_theor_H}. A long-range attraction
of vortices was observed in this interval, leading to different patterns
of intercalated Meissner and Abrikosov-lattice domains
\cite{typeII/1_expA}, often referred to as the type-II/1 state. Recently,
similar vortex configurations have been reported for ${\rm
MgB}_2$~\cite{mosch}, where several carrier bands contribute to the
superconducting condensate. This led to a controversial idea of a special
superconducting type for multiband compounds, named ``type-1.5'', where
multiband materials are considered to comprise several coupled
condensates, one for each band, and a new type arises for at least one
band-condensate being type I and others type II~\cite{mosch,baba}. It was
however immediately argued that unusual features of the ``type 1.5" are
not specific to multiband materials and are actually similar to the
earlier discussed type-II/1 \cite{rev_brandt}. Still, no explanation has
been given how to reconcile the large GL parameter of ${\rm MgB}_2$
($\kappa > 5$, see e.g. Ref.~\cite{zehet}) with the narrow critical range
around $\kappa_0$ where the type-II/1 behavior is expected. To date, the
applicability of the standard classification of superconductors to
multiband materials remained an unresolved and highly debated issue.

We here demonstrate that the appearance of the critical interval near
$\kappa_0$ actually defines a new class of superconducting materials,
which generalizes types II/1 and ``1.5'' and by far exceeds them in
complexity. The new properties of this class are directly related to the
fact that {\it all states of a superconductor, with arbitrary vortex
configurations, have the same energy} at the Bogomolnyi point $(\kappa,T)
=(\kappa_0,T_c)$ when the applied magnetic field is equal to the
thermodynamic critical field $H_c$~\cite{bogomol1,bogomol2,luk}. This
topological degeneracy, lifted at $T < T_c$, is then the source of a wide
diversity of unconventional phenomena in the critical interval of $\kappa$
around $\kappa_0$, where the long-range attraction of vortices mentioned
above is only a particular example. Since this distinct behavior stems
from the critical character of the Bogomolnyi point we refer to it as {\it
critical type} of superconductivity. Interestingly, even some seemingly
understood superconductors (such as Pb) may turn out to belong to the
critical type, so that parts of the long established knowledge in the
field must be revisited. More importantly, we find that coexistence of
multiple bands in a single material gives rise to a large, even giant
enlargement of the critical domain near $\kappa_0$. This appreciably
widens the class of critical superconductors, arguably including most of
the recently discussed materials (such as metal-borides and iron-based
ones~\cite{mult}).

Our work is based on the analysis of the critical GL parameter
$\kappa^\ast$ at which superconductivity should change its type. There are
several standard ways to calculate $\kappa^\ast$: (i) using the condition
of zero surface energy for the superconductor-normal (S-N) interface (the
corresponding critical GL parameter is further denoted as
$\kappa^{\ast}_s$); (ii) from conditions $H_c=H_{c1}$ ($\kappa^{\ast}_1$)
and $H_c=H_{c2}$ ($\kappa^{\ast}_2$), where $H_{c1}$ and $H_{c2}$ are the
lower and upper critical fields, respectively; and (iii) using the
condition of the sign change for the long-range interaction between
vortices~\cite{typeII/1_theor_J,luk} ($\kappa_{li}^\ast$). The GL theory
yields $\kappa^\ast =\kappa_0$ for all above definitions, resulting in
just two possibilities: type I for $\kappa < \kappa_0$ and type II when
$\kappa > \kappa_0$. However, beyond the standard GL theory one obtains
different and temperature dependent values for
$\kappa^\ast$~\cite{typeII/1_theor_J, luk}. Consequences of this fact were
never fully understood and, as a result, are completely ignored in
standard textbooks.

We obtain $\kappa^\ast$ from the Extended GL (EGL) approach, derived by
the perturbation expansion of the microscopic BCS theory in powers of
$\tau=1-T/T_c$ to one order beyond the standard GL theory~\cite{extGL}.
The EGL theory therefore improves the standard GL model but still allows
for simple analytical solutions in many important cases. In order to study
the critical domain around $\kappa_0$, we combine EGL with the expansion
over $\delta\kappa=\kappa-\kappa_0$, in the similar way as was done in
Ref.~\cite{luk}. Then, the density of the Gibbs free energy of the
superconducting state (measured from that of the normal metal) is obtained
at thermodynamic critical field $H=H_c$ as
\begin{align}
g = g^{(0)} +(\delta g^{(0)}/\delta\kappa)\delta\kappa + g^{(1)}\tau,
\label{eq:g_exp}
\end{align}
where the coefficients, calculated at $\kappa=\kappa_0$, are
\begin{subequations}
\label{eq:EGL_energy}
\begin{align}
\label{eq:g_0}
 &g^{(0)}= \frac{1}{2}\Big(\frac{B}{\kappa\sqrt{2}} -
1\Big)^2-|\psi|^2
+ \frac{1}{2}|\psi|^4+ \frac{1}{2\kappa^2}|{\bf D}\psi|^2, \\
&  g^{(1)}=\Big(\frac{B}{\kappa\sqrt{2}}-1\Big)\Big[\frac{1}{2}+ \tilde{c}
+\tilde{G}\big(\tilde{\alpha}-\tilde{\beta}\big)^2\Big]
+\frac{1}{\kappa^2}|{\bf D} \psi|^2 \notag\\
& -
\frac{1}{2}|\psi|^2 +|\psi|^4 +\tilde{c}|\psi|^6 + \tilde{G}|\psi|^2
 \big(\tilde{\alpha}-
\tilde{\beta}|\psi|^2\big)^2
 \notag\\
&+\frac{\tilde{\cal Q}}{4\kappa^4}
\Big[|{\bf D}^2\psi|^2+\frac{1}{3}({\rm rot}{\bf B})^2+{\bf B}^2|\psi|^2\Big]
\notag\\
&+\frac{\tilde{\cal L}}{4\kappa^2} \Big[ 8 |\psi|^2
|{\bf D} \psi|^2 + \big(\psi^{\ast}\big)^2 ({\bf D}
\psi)^2+\psi^2 ({\bf D}^* \psi^*)^2\Big]. \label{eq:g_1}
\end{align}
\end{subequations}
Here ${\bf D}= \nabla + i{\bf A}$, the magnetic field is assumed to be in
the $z$-direction, ${\bf B} = (0,0,B)$, and we used the dimensionless
quantities ${\bf r}/\lambda \sqrt{2}$, $\kappa {\bf A}/ \lambda H_c$,
$\kappa \sqrt{2} {\bf B}/H_c$, $\sqrt{b} \psi/\sqrt{-a\tau}$ and $4\pi
g/H^2_c$, with $H_c=\tau\sqrt{4\pi a^2/b}$ (the lowest order contribution
to the thermodynamic critical field). The order parameter $\psi$ and the
magnetic field $B$~(both in the lowest order in $\tau$) obey the standard
GL equations
\begin{align}
&\psi\big(1-|\psi|^2\big)+\frac{1}{2\kappa^2}{\bf D}^2\psi=0,~ {\rm rot}
{\bf B} = {\bf j}, \label{eq:GL}
\end{align}
where  ${\bf j}= -i(\psi{\bf D}^\ast\psi^\ast - \psi^\ast{\bf D} \psi)$.
Eqs. (\ref{eq:g_exp}) - (\ref{eq:GL}) are valid for a system with an
arbitrary number of carrier bands (see Ref.~\cite{extGL}), where $\psi$ is
a single Landau order parameter \cite{note}. Differences between the band
condensates are accounted for in the coefficients of the equations. In
particular, for the two-band case we obtain
\begin{align}
&a=\frac{a_1}{S} + Sa_2,~b= \frac{b_1}{S^2} + S^2b_2,~{\cal K}=
\frac{{\cal K}_1}{S} +S{\cal K}_2,\notag\\
&\alpha=\frac{a_1}{S} - Sa_2,~\beta = \frac{b_1}{S^2}-S^2b_2,
~\varGamma = \frac{{\cal K}_1}{S}-S{\cal K}_2,\notag\\
&c=\frac{c_1}{S^3} +S^3 c_2,~{\cal Q}=\frac{{\cal Q}_1}{S}
+ S{\cal Q}_2,\;{\cal L} =\frac{{\cal L}_1}{S^2}  + S^2 {\cal L}_2,\notag\\
&\tilde c=\frac{c a }{3 b^2},~\tilde {\cal Q}=\frac{ {\cal Q} a}{{\cal
K}^2},~ \tilde{\cal L}=\frac{{\cal L}a}{b{\cal K}},~\tilde{G} =
\frac{G a }{4g_{12}},\notag\\
&\tilde\alpha=\frac{\alpha}{a}-\frac{{\Gamma}}{{\cal K}},~\tilde\beta =
\frac{\beta}{b}-\frac{{\Gamma}}{{\cal K}}. \label{eq:constants}
\end{align}
where indices label the bands, and $G=g_{11}g_{22} - g_{12}^2$, with
$g_{ij}$ denoting the interaction matrix. The relative weights of the band
contributions to Eqs.~(\ref{eq:constants}) are controlled by the quantity
\begin{equation}
S=\frac{1}{2\lambda_{12}}\!\left[\lambda_{22}-\frac{\lambda_{11}}{\chi} +
\sqrt{\Big(\lambda_{22}-\frac{\lambda_{11}}{\chi}\Big)^2\!\!
+4\frac{\lambda^2_{12}}{\chi}} \;\right], \label{eq:S}
\end{equation}
where $\chi = N_2(0)/N_1(0)$ is the ratio of the band dependent densities
of states (DOS), and $\lambda_{ij}=g_{ij}N(0)$ is the dimensionless
coupling constant, with $N(0)= \sum_jN_j(0)$ the total DOS. Parameters
$a_i,\,b_i, \,c_i,\,{\cal K}_i,\,{\cal Q}_i$, and ${\cal L}_i$ are
calculated separately for each band with a chosen model for the carrier
states (for procedure, see \cite{extGL}). Notice that in the limit $\chi
\to \infty$ or $\chi \to 0$ one recovers the single-band result.

Derivative $\delta g^{(0)}/\delta\kappa$ in Eq.~(\ref{eq:g_exp}) accounts
for both explicit $\kappa$-dependence of the functional and for the
implicit one through the solution $\psi$. Since $\psi$ satisfies
Eq.~(\ref{eq:GL}), i.e. $\delta g^{(0)}/\delta\psi=0$, the implicit
dependence does not contribute to the derivative and one obtains
\begin{align}
\frac{\delta g^{(0)}}{\delta \kappa} = \frac{\partial
g^{(0)}}{\partial\kappa}=-\frac{B}{\kappa^2\sqrt{2}}
\Big(\frac{B}{\kappa\sqrt{2}}-1\Big)-\frac{1}{\kappa^3} |{\bf D}\psi|^2.
\label{eq:dg_0}
\end{align}
Special character of the Bogomolnyi point (BP) follows formally from the
fact that $\psi$ satisfies the Sarma equation $\Pi^-\psi=0$ where $\Pi^\pm
= D_x \pm i D_y$, which leads to relation $|\psi|^2=1-B$ ~\cite{degen}. In
string theories these relations are often referred to as the self-duality
Bogomolnyi equations~\cite{weinberg}. Substituting them into
Eq.~(\ref{eq:g_exp}) and integrating the result, we obtain the total
energy difference as (${\cal G}=\int g~d^3r$)
\begin{align}
{\cal G}=&-\sqrt{2}I\delta\kappa + \tau\Big\{I\big[1-\tilde{c}
+2\tilde{\cal Q}+\tilde{G}\tilde{\beta}(2\tilde{\alpha}
-\tilde{\beta})\big]\notag \\
& + J\big[2\tilde{\cal L}-\tilde{c}-\frac{5}{3}\tilde{\cal Q}
-\tilde{G}\tilde{\beta}^2\big]\Big\}, \label{eq:functional_exp_final}
\end{align}
where $I=\int d^3r|\psi|^2(1-|\psi|^2)$ and  $J=\int d^3r|\psi|^4(1-
|\psi|^2)$, and $g^{(0)}=0$ manifests the BP degeneracy. In the
single-band case this expression is similar to the perturbation functional
used in Ref.~\cite{luk}. The contribution with $\tilde{G}$ in
Eq.~(\ref{eq:functional_exp_final}) appears due to the difference in the
spatial profiles (and, respectively, length-scales) of different band
condensates: $\tilde{G}=0$ for single-band systems.

Equation~(\ref{eq:functional_exp_final}) enables calculating $\kappa^\ast$
according to the definitions given above. This yields the general
expression
\begin{align}
\kappa^\ast = &\kappa_0\Big\{1+\tau\Big[1-\tilde{c}+2\tilde{\cal Q}+
\tilde{G}\tilde{\beta} (2\tilde{\alpha}-\tilde{\beta})\notag \\
& + \frac{J}{I}\big(2\tilde{\cal L}-\tilde{c}-\frac{5}{3}\tilde{\cal
Q}-\tilde{G}\tilde{\beta}^2\big)\Big]\Big\}. \label{eq:kappa_ast}
\end{align}
For $\kappa_s^\ast$, at which ${\cal G}$ calculated for the S-N interface
is zero, the numerical solution yields $J/I=0.559$. For $\kappa_1^\ast$,
obtained from the condition $H_c=H_{c1}$ (which marks the onset of the
stability of an Abrikosov vortex), one obtains $J/I=0.735$. For
$\kappa_{li}^\ast$, at which the long-range asymptote of the vortex-vortex
interaction changes sign, one finds the exact result $J/I=2$. For the
remaining critical parameter $\kappa_2^\ast$ calculated from $H_c=H_{c2}$
one obtains $J/I=0$, as $|\psi| \to 0$.

\begin{figure}[t]
\begin{center}
\resizebox{1.0\columnwidth}{!}{\rotatebox{-00}{\includegraphics{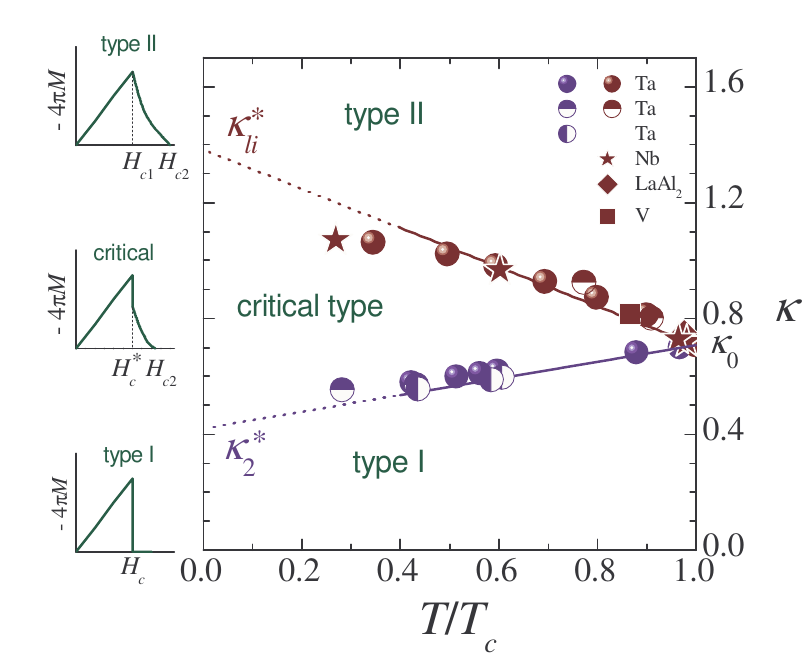}}}
\caption{(Color online) Domain of the critical type for single-band
superconductors. The lines $\kappa_{li}^\ast$ and $\kappa_2^\ast$ are
calculated from the EGL theory [Eq. (\ref{kappa_ast_sb})] and are
material-independent. Symbols represent experimentally obtained boundaries
of the critical domain~\cite{typeII/1_expB,typeII/1_expC,materials} for
several materials. Different shapes of the magnetization curve associated
with different types of superconductivity are drawn schematically in the
left panels.}\label{fig1}
\end{center}
\end{figure}

For a single-band system in the clean limit and with a spherical Fermi
surface one then obtains the following {\it universal expressions},
independent of the particular material parameters:
\begin{eqnarray}
\kappa_{li}^\ast=\kappa_0
(1+0.951\tau),\kappa_1^\ast = \kappa_0(1+0.093\tau),\nonumber\\
\kappa_s^\ast =
\kappa_0(1-0.027\tau),\kappa_2^\ast=\kappa_0 (1- 0.407\tau).
\label{kappa_ast_sb}
\end{eqnarray}
We note that similar linear dependencies were obtained earlier in
Ref.~\cite{typeII/1_theor_J} (within the approximate Neumann-Tewordt
model) and recently in Ref.~\cite{luk} (although with different numerical
coefficients). At $T < T_c$ ($\tau >0$) we obtain $\kappa_2^\ast <
\kappa_s^\ast < \kappa_1^\ast < \kappa_{li}^\ast$. The upper and lower
limits of this inequality define the critical interval $[\kappa_2^\ast,
\kappa_{li}^\ast]$, where superconductivity types I and II interchange. At
$\kappa < \kappa_2^\ast$ only the Meissner state is possible (type I), at
$\kappa > \kappa_{li}^\ast$ the system can develop a mixed state with
repulsive Abrikosov vortices (type II). Experimentally, $\kappa_2^\ast$
and $\kappa_{li}^\ast$ can be obtained from the field dependence of the
magnetization $M(H)$ \cite{typeII/1_expB,typeII/1_expC}, since inside the
critical domain the magnetization has a discontinuity at $H_c^\ast$ ($<
H_c$) \cite{note1}, accompanied by a nonzero tail at $H_c^\ast < H  <
H_{c2}$ due to the existence of a mixed state. Fig.~\ref{fig1} shows
$\kappa_{li}^\ast$ and $\kappa_{2}^\ast$ from Eq.~(\ref{kappa_ast_sb})
alongside experimental data obtained for nearly clean single-band
materials~\cite{typeII/1_expB,typeII/1_expC}. The experimental boundaries
of the critical interval are indeed material-independent and follow
straight lines down to temperature $T \approx 0.4 T_c$, in remarkable
agreement with our theoretical prediction.

\begin{figure}[t]
\resizebox{1.0\columnwidth}{!}{\rotatebox{0}{\includegraphics{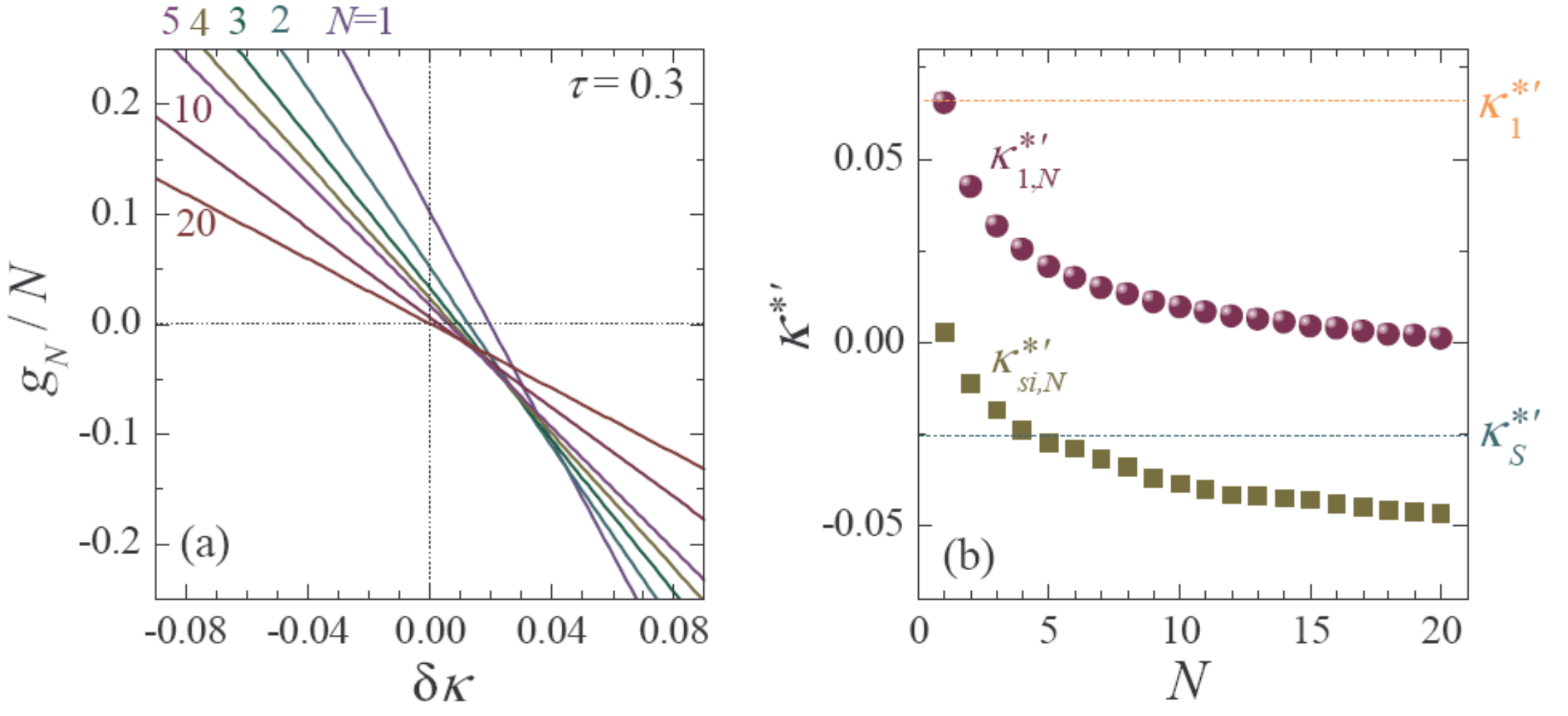}}}
\caption{(Color online) Energy and interaction of multi-quanta vortices in
the critical domain: (a) the Gibbs free energy (difference to the normal
state) of an isolated $N$-quanta vortex as a function of $\delta\kappa
=\kappa-\kappa_0$, calculated at $T=0.7T_c$ ($\tau=0.3$) and $H=H_c$, and
normalized by $N$. (b) Tangents ($d\kappa^\ast/d\tau$) of
$\kappa^\ast_{1,N}$ (where $N$-quanta vortices are stabilized) and
$\kappa^\ast_{si,N}$ (where short-range interaction of $N$-quanta vortices
changes sign) as a function of $N$. Dashed lines show the slopes of
$\kappa^\ast_1$ and $\kappa^\ast_{s}$ for comparison.} \label{fig2}
\end{figure}

Although differences between $\kappa^\ast$ obtained using different
definitions have been noted earlier~\cite{typeII/1_theor_J}, their
relation to the Bogomolnyi point has been noted only recently~\cite{luk}.
However, the nontrivial internal structure of the critical domain, related
to the topological degeneracy of the Bogomolnyi point, has not been
considered. When this degeneracy is lifted at $T < T_c$, different
$\kappa^\ast$ delimit subdomains of very different magnetic
behavior~\cite{note2}. Here we describe the critical-domain structure by
addressing crucial changes in the vortex behavior at $\kappa_1^\ast$ and
$\kappa_s^\ast$.

When $\kappa_1^\ast < \kappa < \kappa_{li}^\ast$, the system hosts
single-quantum vortices for $H > H_{c1}$, in an Abrikosov lattice.
However, unlike vortices in conventional type-II superconductors, these
vortices are attractive at long distances~\cite{typeII/1_theor_J,moh}, so
the lattice constant has a finite maximal value. One consequence of this
is a first-order transition with a magnetization jump at $H_c^\ast=H_{c1}$
(see Fig.~\ref{fig1}). This behavior is known in literature and often
referred to as type II/1~\cite{typeII/1_expB,typeII/1_theor_J,
type_II/1_theor_H, rev_brandt}. However, so far it was associated with the
entire critical domain, i.e., for $\kappa_2^\ast < \kappa <
\kappa_{li}^\ast$, and this is not correct.

Namely, at $\kappa < \kappa_1^\ast$, an isolated single-quantum vortex
becomes energetically unfavorable at $H_c^\ast$, since $H_c^\ast < H_c$
and $H_{c1} > H_c$. However, we still have the mixed state, since $H_c <
H_{c2}$. As the closest alternative, we have put into consideration the
$N$-quanta vortices, even though they are supposed to be unstable in the
conventional type-II picture. Quite surprisingly, Fig.~\ref{fig2}(a) shows
that when $\kappa$ decreases in the interval $[\kappa_s^\ast,
\kappa_1^\ast]$, the multi-quanta vortices proliferate, i.e., vortices of
progressively larger $N$ become more favorable than $N$ isolated
single-quantum vortices. As a consequence, the Meissner state becomes
unstable at the field $H_c^\ast =H_{c1,N} < H_c < H_{c1}$, where
$H_{c1,N}$ labels $H_{c1}$ for the $N$-quanta vortex.

Using the condition $H_c=H_{c1,N}$, we define the critical parameter
$\kappa_{1,N}^\ast$ at which $N$-quanta vortices first appear in the
system. We also consider $\kappa_{si,N}^\ast$ and $\kappa_{li,N}^\ast$ at
which the short-range and, respectively, the long-range asymptotes of the
vortex-vortex interaction change their sign. In Fig.~\ref{fig2}(b), we
show the calculated $\kappa_{1,N}^\ast$ and $\kappa_{si,N}^{\ast}$, which
also satisfy Eq. (\ref{eq:kappa_ast}) but for $N$-dependent integrals $J$
and $I$. $\kappa_{si,N}^{\ast}$ decreases monotonically for increasing
$N$, leading to a conclusion that only vortices with progressively larger
$N$ remain stable against fusion when decreasing $\kappa$. On the other
hand, $\kappa_{li,N}^\ast=\kappa_{li}^\ast$ is independent of $N$.

\begin{figure}[t]
\begin{center}
\resizebox{0.8\columnwidth}{!}{\rotatebox{-00}{\includegraphics{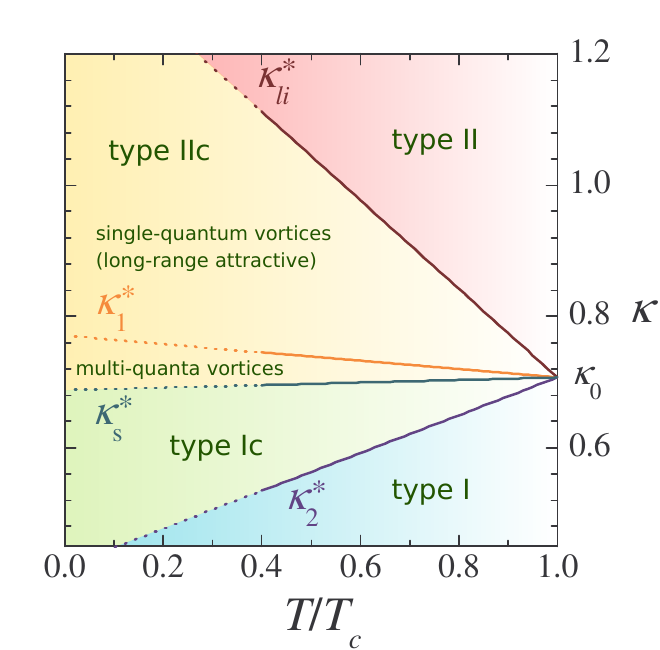}}}
\caption{(Color online) Internal structure of the critical domain,
delimited by the line $\kappa_s^\ast(T)$ which separates subtypes Ic (with
mixed state, but without isolated vortices) and IIc (the mixed state with
single-quantum or multi-quanta vortices, attractive at long distances;
multi-quanta vortices proliferate below the line
$\kappa_1^\ast(T)$).}\label{fig3}
\end{center}
\end{figure}

When $\kappa$ descends below $\kappa^\ast_s$, the S-N surface energy
becomes positive and fragmentation of S-N interfaces into vortices is no
longer energetically advantageous. Isolated vortices are, therefore,
unstable for any $N$ and $H_c^\ast\not=H_{c1,N}$. Nevertheless, as $H_c <
H_{c2}$, the mixed state must exist and should be densely inhomogeneous.
An educated guess into its shape (e.g., a dense distribution of strongly
overlapping multiple-quanta vortices, stripe structures, or something
similar to the Fulde-Ferrel-Larkin-Ovchinnikov pattern~\cite{FFLO}) is
necessary to determine the corresponding $H_c^\ast$. For the present
analysis, it is most important that the single-vortex picture becomes
inadequate here. We label this subdomain critical type I (type Ic). In
contrast, when $\kappa >\kappa_s^\ast$, isolated vortices are more
energetically favorable than the Meissner state at $H_c^\ast < H <
H_{c2}$. However, unlike the conventional type II, vortices are attractive
at long distances and each can carry multiple flux-quanta. We refer to
this behavior as critical type II (type IIc). The summary of this analysis
is illustrated in Fig. \ref{fig3}, with delimiting lines obtained from
Eq.~(\ref{kappa_ast_sb}). It must be noted at this point that the complete
analysis should include the field dependence of the most favorable
configurations, since our present consideration of possible intermediate
states is restricted to fields in vicinity of $H_c$.

\begin{figure}[t]
\begin{center}
\resizebox{1.0\columnwidth}{!}{\rotatebox{0}{\includegraphics{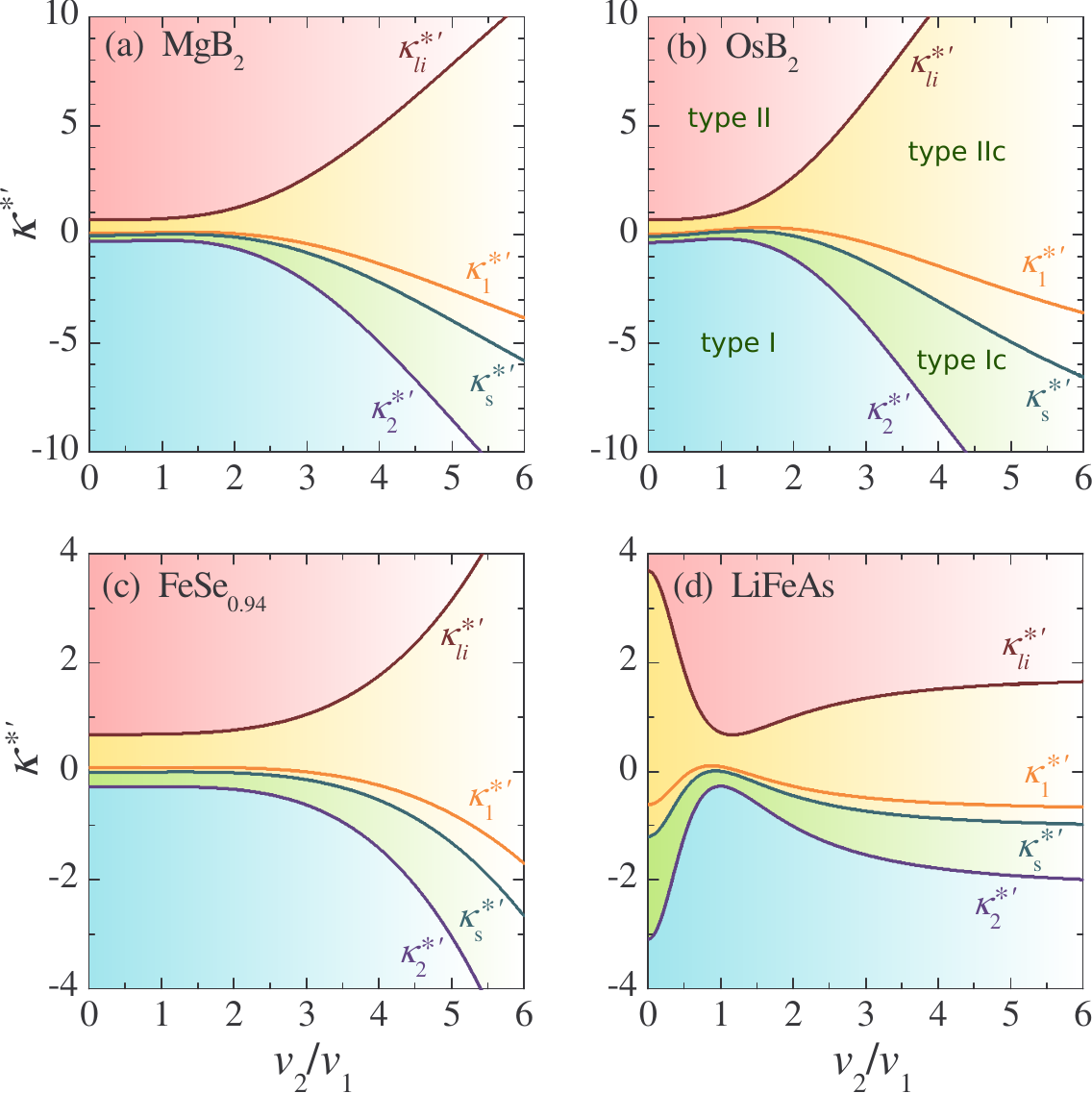}}}
\caption{(Color online) Widening of the critical domain for two-band
superconductors: growing tangents ($\kappa^{\ast\prime}=
d\kappa^\ast/d\tau$) of critical parameters $\kappa^\ast_{li}$,
$\kappa^\ast_1$, $\kappa^\ast_s$ and $\kappa^\ast_2$ as a function of the
ratio $v_2/v_1$ of the band Fermi velocities, calculated from the EGL
theory for the available material parameters of ${\rm MgB}_2$~(a), ${\rm
OsB}_2$~(b), ${\rm FeSe_{0.94}}$~(c) and ${\rm FeSe_{0.94}}$~(d).}
\label{fig4}
\end{center}
\end{figure}

As best seen from Fig.~\ref{fig3}, in single-band superconductors the
critical domain is limited to GL parameters relatively close to
$\kappa_0$, thus critical superconductivity is of relevance to a
restricted number of materials. This changes drastically for multiband
superconductors, where the interband coupling enhances non-local
effects~\cite{note3}, known to be responsible for the long-range vortex
attraction~\cite{rev_brandt}. An amplification of the critical type of
behavior can thus be anticipated. Our calculations for the two-band system
fully confirm this expectation. Fig.~\ref{fig4} demonstrates a very
significant widening of the critical region for two-band superconductors
as a function of the ratio of the band Fermi velocities $v_2/v_1$. The
conclusion that critical superconductivity is enhanced by disparate
length-scales of two or more involved band-condensates is rather general
and holds for a wide selection of the material parameters. In particular
the results in Fig.~\ref{fig4} are calculated for the parameters of:
(a)~${\rm MgB}_2$~\cite{MgB2}, $\lambda_{11} = 1.88$, $\lambda_{22} =
0.5$, $\lambda_{12} = 0.21$, and $N_2(0)/N_1(0)=1.33$); (b)~${\rm
OsB}_2$~\cite{OsB2}, $\lambda_{11} = 0.387$, $\lambda_{22} = 0.291$,
$\lambda_{12} = 0.0084$, and $N_2(0) /N_1(0)=1.22$, (c)~${\rm
FeSe_{0.94}}$~\cite{FeSe}, $\lambda_{11} = 0.48$, $\lambda_{22} = 0.39$,
$\lambda_{12} = 0.005$, and $N_2(0)/N_1(0) = 1$, and (d)~${\rm
LiFeAs}$~\cite{LiFeAs} $\lambda_{11} = 0.63$, $\lambda_{22} = 0.64$,
$\lambda_{12} = 0.008$, and $N_2(0)/N_1(0)=1$. As a main conclusion of
Fig.~\ref{fig4}, the critical domain in two band materials is always
larger than in single-band ones. One should however notice its
extraordinary sensitivity to $v_2/v_1$: for $v_2 > 2v_1$ in
Figs.~\ref{fig4}(a)-(c) and for $2v_2 < v_1$ in Fig.~\ref{fig4}(d),
tangents of $\kappa_{li}^\ast$ and $\kappa_2^\ast$ change by an order of
magnitude or even more. As a consequence, critical type of
superconductivity becomes very likely at low temperatures and should
always be reckoned with in multiband materials.

In summary, we have shown analytic distinction of the class of critical
superconductors, which draw their unique properties from the physics of
the Bogomolnyi critical point. Critical type of superconductivity is
intrinsic to some long known superconductors, such as e.g. Nb, or Ta with
impurities, but could also be relevant to Pb which was always assumed
type-I \cite{note4}. Our analysis suggests that critical superconductivity
should always be taken into account in the studies of novel
superconducting compounds, particularly multiband ones with notably
different Fermi surfaces (such as e.g. ${\rm FeSe}_x{\rm
Te}_{1-x}$~\cite{mult}). In critical superconductors, the rich structure
of the critical domain will undoubtedly reflect itself in a wealth of
unconventional superconducting states, with unique sensitivity to the
applied field, current, temperature but also to different geometric
parameters of a specimen. Detailed description of those states warrants
further theoretical as well as experimental investigations. Finally, it is
important to note that in string models mathematically similar to the
superconducting theory (such as the Abelian Higgs model
\cite{bogomol1,weinberg}), by analogy to our findings a similar critical
domain of unusual equilibria could also be observed.

This work was supported by the Flemish Science Foundation (FWO-Vl), and
the Methusalem program. A.A.S. acknowledges support of Brazilian agencies
CNPq and FACEPE (APQ-0589-1.05/08). The authors express their gratitude to
I. Luk'yanchuk for reading the manuscript and important comments.



\begin{thebibliography}{99}
\bibitem{note} Strictly speaking, the single Landau order parameter $\psi$
appears in the theory only in the absence of an additional symmetry
between the contributing bands. This interesting analysis is beyond the
scope of the present Letter and will be considered elsewhere (see the
basic ideas in Ref.~\cite{orl}).
\bibitem{note1} $H_c^\ast$ can not exceed $H_c$ as follows from
$\int_0^{H_{c2}} M(H)\,dH=H^2_c/(8\pi)$.
\bibitem{materials} In experiments with ${\rm Ta}$ and ${\rm Nb}$ parameter
$\kappa$ is varied by the nitrogen doping. Fig.~\ref{fig1}(a) summarizes
experimental data for samples with the different nitrogen concentration.
In all specimen the electron mean-free path remains smaller than the BCS
coherence length.
\bibitem{note2} It should be noted that critical parameters $\kappa^\ast$
discussed so far do not exhaust all possibilities: any topologically
different solution of Eqs.~(\ref{eq:GL}) can be associated with its own
critical parameter $\kappa^\ast$ defined at the onset of stability of that
particular solution. This leads to further, but less fundamental, possible
subdivisions of the critical domain.
\bibitem{note3} Nonlocality is generally related to the spatial gradients
that are larger for systems with several coupled components.
Mathematically greater non-local effects in a system of coupled equations
follows from the increase in the effective order of the derivatives
(spatial memory effect).
\end{thebibliography}
\end{document}